\documentclass{llncs}

\usepackage[english]{babel}
\usepackage[utf8]{inputenc}
\usepackage{latexsym,amssymb,amstext,amsfonts,amsmath,graphicx,setspace,mathtools,bm}
\usepackage[colorinlistoftodos]{todonotes}
\usepackage{enumerate} 
\usepackage{paralist}
\usepackage{float}
\usepackage{array}
\usepackage{comment}
\usepackage{pdflscape}
\usepackage{rotating}

\usepackage{tikz}
\usetikzlibrary{fit,positioning}
\usetikzlibrary{calc}

\newcommand{\BlueComment}[1]{\color{blue}{#1}\color{black}}

\newcommand*\dif{\mathop{}\!\mathrm{d}}
\DeclareMathOperator*{\E}{\mathbb{E}}

\begin{document}

\title{Measures of Tractography Convergence}
\titlerunning{Tractography Kernel Surprise}  % abbreviated title (for running head)
%                                     also used for the TOC unless
%                                     \toctitle is used
%
\author{Daniel C. Moyer\inst{1}\inst{2} \and Paul M. Thompson\inst{1} \and Greg ver Steeg\inst{2} }

\authorrunning{Daniel Moyer et al.} % abbreviated author list (for running head)

%%% list of authors for the TOC (use if author list has to be modified)
\tocauthor{Daniel C. Moyer, , and }

\institute{Imaging Genetics Center, Stevens Institute for Neuroimaging and Informatics, University of Southern California, Los Angeles, CA 90007, USA
\and
Information Sciences Institute, Marina del Rey, CA 90292, USA\\
\email{moyerd@usc.edu}}

%\author{Anon}
%\institute{***}

\maketitle              % typeset the title of the contribution

\begin{abstract}
In the present work, we use information theory to understand the
empirical convergence rate of tractography, a widely-used approach to reconstruct anatomical fiber pathways in the living brain.
Based on diffusion MRI data, tractography is the starting point for many methods to study brain connectivity.
Of the available methods to perform tractography, most reconstruct a finite set of streamlines, or 3D curves, representing probable connections between anatomical regions, yet relatively little is known about how the sampling of this set of streamlines affects downstream results, and how exhaustive the sampling should be.
Here we provide a method to measure the information theoretic surprise (self-cross entropy) for tract sampling schema. We then empirically assess four streamline methods.
We demonstrate that the relative information gain is very low after a moderate number of streamlines have been generated for each tested method. The results give rise to several guidelines for optimal sampling in brain connectivity analyses. 
\end{abstract}

\section{Introduction}
\label{sec:intro}

 It is common practice in brain imaging studies to estimate the trajectories of white matter fascicles, the bundles of axons that connect different gray matter regions in the brain. It is difficult to observe the fascicles directly \textit{in vivo}; instead, researchers currently rely on diffusion imaging - a variant of standard anatomical MRI - and the ``tube like'' structure of most major neural pathways to reconstruct the brain's major fiber bundles. We observe measurements of water diffusion in brain tissue, fit local diffusion models, and then reconstruct the trajectories by curve fitting. This process is known as tractography, and reconstructed fibers are referred to as tracts\footnote{It is important to distinguish between white matter fibers (fascicles) and observed ``tracts.'' In this paper we use ``tracts'' to denote the 3D-curves recovered from diffusion-weighted imaging via tractography algorithms.}.

The number of possible curves is combinatorially large, and grows rapidly as image resolution increases. Closed form likelihoods for sets of tracts (tractograms) have not yet been found even under simple local models. Thus, tract sampling is performed, where only a subset of the possible tracts are recovered. Sampling is usually performed in an ad-hoc manner, and post-hoc analyses usually treat tracts as direct observations instead of realizations of global trajectories from local models. The concept that tracts are drawn from a distribution is rarely acknowledged, and the issue of convergence is almost never addressed.

A simple question remains unanswered: how many tracts should we sample? Even in the most general sense there is no consensus on this topic. Recent works have used 100,000 \cite{girard2014towards}, 500,000 \cite{garyfallidis2017recognition} and 10,000,000 \cite{smith2015sift2} tracts for scans of isotropic resolution 2 mm (with 64 directions), 1.25 mm (90 directions), and 2.5 mm (60 directions), respectively. The pragmatic tractography user might then think to simply take the maximum of these options to ensure convergence; for small studies this may be feasible, but given the trajectory of imaging studies - now performed in biobanks of over 10,000 scans - this choice clearly will not scale to the regimes demanded by current and future clinical studies\footnote{The largest study to date aims to have 100,000 subjects participating in the imaging cohort \cite{miller2016multimodal}. By our back-of-the-envelope estimate, 100,000 subjects with 10,000,000 tracts each would require about 1.5 petabytes of disk space, just for the tractograms.} using current technology. The answer to whether or not such large tractogram sizes are required will dictate practical planning in large studies. %If such tractography sample sizes are necessary, then there is strong impetus to develop new tractography representation and compression methods; if they are not necessary, both computational and human resources should be allocated elsewhere.

It is useful then to understand empirically the sample complexity of these phenomena. If we choose a sample size too small our tractograms will be sensitive random fluctuations; if instead we choose a sample size too large, we could be wasting very large amounts of storage and computation. The trade-off between accuracy and cost forms a curve of solutions to the (ill-posed) question of ``how many tracts?'' In the present work we provide a method for measuring this trade-off. We leverage cross entropy as a relative measure of accuracy, applying our estimator to a test set of subjects under several different tractography methods.
%related trade-off curve between the number of samples and accurately describing the differences between distributions (e.g. between two tracking methods).

%We first describe a conceptual framework, clarifying theoretical properties of tractography (opposed to e.g. direct observation). Based on this framework we then describe our method and present our empirical results. It is paramount that these results be taken in context. Different choices of local diffusion models, tracking parameters, seeding criteria, etc. will change the distribution of tracts and thus may change the convergence rate of any estimator of that distribution. \BlueComment{Instead of using our empirical results as given for different circumstances, we provide code that may be used to produce similar results in a user's context, using their own choices of parameters. }

\subsection{Relevant Prior Work}

Surprisingly literature on the number of samples required is sparse. In the documentation of Probtrackx \cite{behrens2007probabilistic} the number of posterior samples is taken to be large (5,000 per vertex) specifically to ensure convergence. This solution is not feasible for most other methods, as Probtrackx does not usually keep computed trajectories, but is the most direct reference to tractogram convergence that we found in the literature.

In Cheng et al. \cite{cheng2012optimization} the authors address the convergence of network measures. While this is a similar issue, it is seen through the lens of graph theory. In a network context the absolute (objective) number of tracts strongly influences the downstream measures. Thus, the authors advise \emph{against} larger numbers of seeds in order to avoid spurious fibers. Similarly, Prasad et al. \cite{prasad2013tractography} investigate effect of the number of streamlines used on a group-wise comparison task. The authors note that larger numbers of streamlines tend to decrease p-values in nodal degree comparisons. 

Similar to this, Maier et al. \cite{maier2017challenge} provide a tractography challenge in which the measure of success depends in part on how few false positives were identified. This work is an important if not critical step for the community to address issues raised by others (e.g., Thomas et al. \cite{thomas2014anatomical}), but its evaluation design runs counter to our premise here that tractography is a simulation. It is equivalent to taking the support of the distribution as its characterizing quality.

Jordan et al. \cite{jordan2018cluster} propose a method for measuring streamline reproducibility. This method is essentially the same as proposed here, using a KDE-like form to assign a weight (density) to each streamline. The authors do not identify this as a KDE (and accordingly use a non-standard power-law kernel), but otherwise produce the same method.

\section{Framework and Methods}
\label{sec:method}

\subsection{Minimal Surprise as a stopping condition}

%To perform streamline tractography is to sample from a distribution of possible tracts through simulation. Diffusion MRI does not directly observe whole white matter fascicles; \BlueComment{the unit of observation is a q-space measurement of diffusion propensity for given gradient angles, which is resolved and processed into quantized voxel measurements}. Researchers simulate fascicle trajectories from fitted local (voxel-wise) models, which have parameters estimated from the diffusion signal (for reference and review, see Sotiropoulos and Zalesky \cite{zalesky2017howwhybut}). Samples of tracts serve as proxies of their underlying distribution, either explicitly via distribution estimation or implicitly as tracts become the object of study.

%The distance between the sample (empirical) distribution and the generating distribution is the

%It is then useful to know how close a sample is to its generating distribution. 

%We would like to know for a given finite sample how representative is that sample of the underlying distribution. 
%From these properties we justify our convergence criterion, information-theoretic surprise.

%Measuring distances between distributions is well studied. 

%To perform streamline tractography is to sample from a distribution of possible tracts through simulation.

Diffusion MRI does not provide direct observations of whole white matter fascicles; the unit of observation is a $q$-space measurement of diffusion propensity for given gradient angles, which is resolved and processed into quantized voxel measurements. Researchers simulate fascicle trajectories from fitted local (voxel-wise) models, which have parameters estimated from the diffusion signal (for reference and review, see Sotiropoulos and Zalesky \cite{zalesky2017howwhybut}). Some of these simulations are deterministic given a seed point, while others are stochastic. In both cases, however, seed points are not bijective with tracts. If the choice of seed point is random, it is then reasonable to view the tractography generation process as sampling from a distribution $P$ defined over the space of possible tracts, stratified by seed points.

The space of possible tracts has at minimum a qualitative notion of similarity. While there is no agreed upon metric, various distances and similarities have been proposed \cite{o2007automatic,wassermann2010unsupervised,garyfallidis2012quickbundles}. Choosing a reasonable metric, a simplistic estimator $\hat{P}$ of tract distributions can be constructed using Kernel Density Estimation (KDE). These estimators enjoy concentration results\footnote{In short, we are asserting that $\hat{P}$ converges as the number of samples used to construct it increases. This does not guarantee that $\hat{P}$ converges to $P$.} even in our general context \cite{coifman2006diffusion} with minimal constraints on metric choice or kernel shape. The construction of distribution estimates $\hat{P}$ lead us to a na{\"i}ve condition for assessing convergence: we should stop sampling when $\hat{P}$ stops gaining information (``stops learning'').

$\hat{P}$ is a representation of our beliefs about $P$. A sample from $P$ containing information not in $\hat{P}$ (i.e., contrary to our beliefs) is then qualitatively surprising. As we update our beliefs $\hat{P}$ we learn about $P$; when the surprise no longer decreases we have stopped updating our beliefs, so we have stopped gaining information, which is exactly our stopping condition. Thus we should sample until we have minimal surprise.

Before moving on, it should be made explicit: clearly $\hat{P}$ will be misspecified. Without a natural metric, it is not clear if an estimator class could ever be guaranteed to include $P$ in every case. With this in mind, our provided method is left open-ended toward choice of metric, kernel shape, smoothing parameter, etc. Further, for different datasets, the convergence rate of $\hat{P}$ will be different. Higher resolution data will provide more information, providing more intrinsic information in $P$, meaning $\hat{P}$ has more surprise to overcome. The point of this paper is not to estimate $P$ exactly, nor to establish objective numerical standards for the number of tracts for every case, but to provide a method to investigate the complexity of tract generation. It is our belief that with reasonable choices a conservative estimate can be made to the number of samples required.

%For an increasing number of samples, qualitatively good estimates $\hat{P}$ have low variation; conclusions drawn from smaller datasets are less robust to sampling effects than larger datasets. It is useful to understanding objectively how many samples are required to avoid such variance.

\subsection{Quantitative Measures of Surprise}

Kullback-Leibler divergence (KL divergence) is a classic measurement for the differences between distributions. While it has numerous interpretations and applications, in the information theory context it measures the ``extra information'' required to encode samples drawn from $P$ using the optimal encoding for $\hat{P}$ \cite{cover2012elements}. This is often used as a pseudo-distance, as ``close'' distributions have similar optimal encodings.

KL divergence has the following analytic form, with $0 \log 0 = 0$ by convention:
\begin{align}
KL[ ~P ~|~ \hat{P}~] & = \E[\log P(x) - \log \hat{P}(x)] = \int P(x) [\log P(x) - \log \hat{P}(x) ]dx\\
& = 
\underbrace{\int P(x) \log\left(P(x)\right) dx }_{\text{Negative Entropy}}   
\underbrace{-\int P(x)\log\left( \hat{P}(x) \right) dx}_{\text{Cross Entropy}}.
\end{align}
For any $P$ the Negative Entropy term is clearly fixed, regardless of $\hat{P}$. Following our previous intuition, this corresponds to the intrinsic surprise of $P$, the number of bits required to encode $P$ under the optimal encoding. The Cross Entropy term corresponds to the induced surprise of using $\hat{P}$'s encoding in place of $P$. It is thus our desired measure of surprise; measuring Cross Entropy is the focus of our empirical method. Both have units of \emph{bits} (or \emph{nats}) of information.

%Surprise is measured in bits, a unit of information (). Continuing the estimation context, we view a sequence of estimates $Q_n$ (fit from an increasing number $n$ samples) as providing 

\subsection{ Fixed-KDE Cross Entropy}

%However, in both situations it is understood that ``rare'' tracts, i.e. the noise that is not often observed, do not characterize the tractography method, but the more likely ones.

%A question then arises: how many streamlines should we sample to have an acceptable depiction of the underlying streamline distribution? Clearly the answer to this question depends on many factors; we seek to construct general criteria that will, given a choice of scan parameters, local diffusion model, tracking method, tracking parameters, etc., produce a trade-off curve between the number of samples and a accuracy index.

We construct a simple estimator of Cross Entropy using Kernel Density Estimation (KDE). Let $\{x_i\}_{i=1}^N$ be samples from a distribution $P$, and let $d(x_i,x_j)$ be a metric. The (unnormalized) squared exponential kernel function for distance $d$ is defined as
\begin{align}
k(x_i,x_j) = \exp(-\gamma d(x_i,x_j)).\label{eq:kern}
\end{align}
Using this function, we can then define an unnormalized Gaussian KDE:
\begin{equation}
\hat{Q}(x_i) = \frac{1}{N} \sum_{j=1}^N k(x_i,x_j)
\end{equation}
We use $Q(x_i)$ to emphasize that this distribution is off by a normalizing constant, and $\hat{Q}(x_i)$ to denote that this is an estimate of ${Q}$ based on $\{x_i\}_{i=1}^{N}$. We assume there exists a finite normalization constant, $Z = \int Q(x) dx < \infty$. An approximation of $P$ the distribution of tracts is then
\begin{align}
\hat{P}(x) & = \frac{1}{Z}\hat{Q}(x) = \frac{1}{ZN} \sum_{j=1}^N k(x,x_j) \label{eq:dist-approx}
\end{align}
Using Eq. \ref{eq:dist-approx}, we can form an approximation to %the information theoretic surprise and
the Cross Entropy up to a constant.
\begin{align}
%\text{Surprise: }~~~~~~~~ \text{Sur}[X; \hat{P}] & = -\log \hat{P}(X) \\
\text{Cross entropy: }~~\text{H}[\hat{P}(x); P(x)] & = - \int P(x) \log\hat{P}(x) \dif x \\
& = - \int P(x) \log\hat{Q}(x) \dif x ~+~ Const \label{eq:ce-const}\\
& \approx -\frac{1}{N}\sum_{x_i} \left[ \log \frac{1}{N}\sum_{j=1}^N k(x_i,x_j)\right] ~+~ Const \label{eq:ce-approx}
\end{align}
%Surprise is the point-wise measurement of cross entropy, and cross entropy is the expected surprise. Given a likelihood function $\hat{P}$ the surprise of an observation $x$ is $-\log\hat{P}(x)$; given the underlying distribution $P$, the expected surprise $\E_P[-\log \hat{P}]$ is the cross entropy.

%We derive surprise directly from the KL Divergence in Appendix \ref{sec:appendix-surprise}; in summary, as $\hat{P}\rightarrow P$ the cross-entropy converges to $-\int P \log P(x)$, which is a constant value. Thus, in order to empirically measure the convergence of $\hat{P}$ as we increase $N$, we can track the value of $\text{H}[\hat{P}(x); P(x)]$, which is equivalent up to a constant to the average log row sum of the kernel matrix. Computing $\text{H}[\hat{P}(x); P(x)]$ requires computing the kernel matrix, and thus is $O(N^2)$. However, it does not require that it be held in memory. Indeed, given $M > N$ samples we could compute the row sums of each of the $M$ points using the first $N$ points, iteratively increasing $N$ and plotting the cross entropy at each step.

%\subsection{Two Distribution Case}

%In many cases we have two separate distributions $P_1$ and $P_2$ which we know \emph{a priori} to have differences, and in which the objective is to detail where the distributions differ. In order to assess how well the differences are understood, we 

\subsection{Calculating Tract Sample Cross Entropy}

Several metric spaces have been proposed \cite{o2007automatic,garyfallidis2012quickbundles} for streamlines. These allow us to apply our KDE estimator to the streamline distributions. We use the MDF distance of Garyfallidis et al. \cite{garyfallidis2012quickbundles} as the distance metric in Eq. \ref{eq:kern}. We then estimate the Cross-Entropy of samples to tract distributions in these spaces using Eq. \ref{eq:ce-approx}. While Eq. \ref{eq:ce-approx} is always off by an unknown constant, the rate of information gain is preserved. This allows us to measure the marginal value (in bits) of increasing sample sizes. Further, for fixed $\gamma$ the objective information content is comparable across distributions.

As tracts are simulated and each observation is independent, many issues are simplified in our approximation. In particular, we separate the $\{x_i\}$ and $\{x_j\}$ samples, letting the number of $x_i$ be large ($\sim 10^6$) no matter the choice of $N$ (the number of $x_j$) in order to ensure a good estimate of the outer expectation. We can further separate the kernel computation step and the summation step, which allows for fast bootstrap estimates as well as faster computation for increasing sizes of $N$.

\subsection{Scale Considerations}
\label{subsec:scale}

Note that each KDE used fixed and un-tuned bandwidth $\frac{1}{\gamma}$. Thus, our KDE converges to $P * K$ instead of $P$, where $K(x) = Z^{-1} k(x,0)$. We believe this to be a feature and not a bug: first, though the tractography simulation is often conducted with high numerical precision (using, e.g., floating point representation), the trajectories of streamlines are determined in part by the diffusion measurements. These measurements are ill-resolved relative to machine precision. Second, later analyses of streamline distributions are rarely conducted at a local scale. Fixing $\gamma$ allows for ``convergence'' to be contextualized to a particular spatial scale, albeit after the choice of a tract metric. Thus, if the practitioner does not need to resolve small scale differences, a corresponding choice of $\gamma$ can be made.

%Third, the KL divergence between the $P$, $P * K$
%\begin{align}
%KL[P | P * K] = H[P] - H
%\end{align}

%\subsection{Streamline Kernels and Kernel Spaces}

%The space of streamlines is complicated. The representation provided by (discretizations of) physical space is a sequence of points of varying length, which is inconvenient for defining probability measures. Alternative representations have been suggested, in particular reproducing kernel Hilbert space (RKHS) based approaches \cite{o2007automatic,wassermann2010unsupervised}.%, which map the na{\"i}ve representation into a high dimensional space implicitly via kernel functions
%These methods define a $k: \mathcal{X} \times \mathcal{X} \rightarrow \mathbb{R}$. 

\section{Empirical Assessment}

We apply the Cross Entropy estimator to 44 test subjects from the publicly available Human Connectome Project dataset \cite{van2013wu}. We used the minimally preprocessed T1-weighted (T1w) and diffusion-weighted (DWI) images rigidly aligned to MNI 152 space. Briefly, the preprocessing of these images included gradient nonlinearity correction (T1w, DWI), motion correction (DWI), eddy current correction (DWI), and linear alignment (T1w, DWI).  We used the HCP Pipeline FreeSurfer \cite{fischl2012freesurfer} protocol, running the recon-all pipeline.

Tractography was conducted using the DWI in 1.25-mm isotropic MNI 152 space.
Three separate types of tractography were conducted, using three separate local models and two different tracking methods. All three were performed using the Dipy \cite{garyfallidis2014dipy} implementation. The local models were as follows:
\begin{enumerate}
\item Constrained Spherical Deconvolution (CSD) \cite{tournier2008resolving} with a harmonic order of 8
\item Constant Solid Angle (CSA) \cite{aganj2010reconstruction} with a harmonic order of 8
\item Diffusion Tensors (DTI) (see, e.g., \cite{le2001diffusion})
\end{enumerate}
Probabilistic streamline tractography was performed using Dipy's probabilistic direction getter for CSD and CSA local models, while deterministic tracking was completed using EuDX \cite{garyfallidis2013towards} for DTI and the Local Tracking module for CSD. For each local model streamlines were then seeded at 2 million random locations (total) in voxels labeled as likely white matter by FSL's FAST \cite{zhang2001segmentation} and propagated in 0.5 mm steps, constrained to at most $30^\circ$ angles between segments. Each set of tracts was split into two equally sized subsets. Cross-entropy was then calculated using Eq. \ref{eq:ce-approx} for five separate length-scale choices $\gamma=\{\frac{1}{4},\frac{1}{2},1,2,4\}$, using 1000 streamline intervals to allow for parallelization and efficient storage. This portion of the procedure was written in C++ with OpenMP\footnote{Please contact the authors to access the code.}.

\begin{sidewaysfigure*}
\centering
\includegraphics[width=19cm]{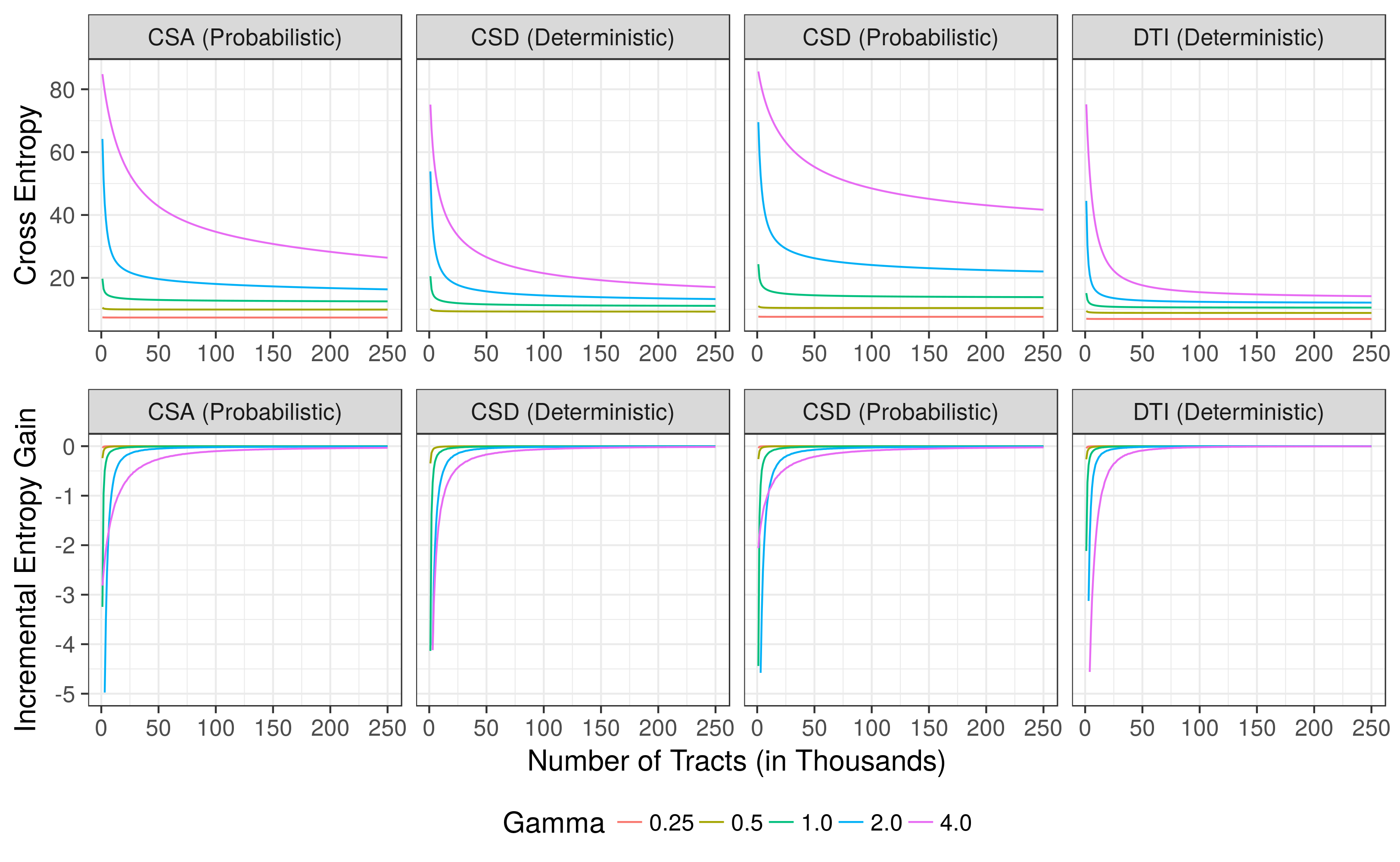}
\caption{\textbf{Top row:} we plot the self cross-entropy (surprise) of each method as a function of the number of tracts, conditioned on the length scale $\gamma$ (given by the plot colors). \textbf{Bottom row: } we plot the incremental (marginal) entropy gain per 1000 streamlines, also conditioned on the length scale. In both row we eschew plotting the x-axis to the full one million samples as the unplotted section is essentially flat.}
\label{fig:ent}
\end{sidewaysfigure*}

\begin{figure}[t]
\centering
\includegraphics[width=\textwidth]{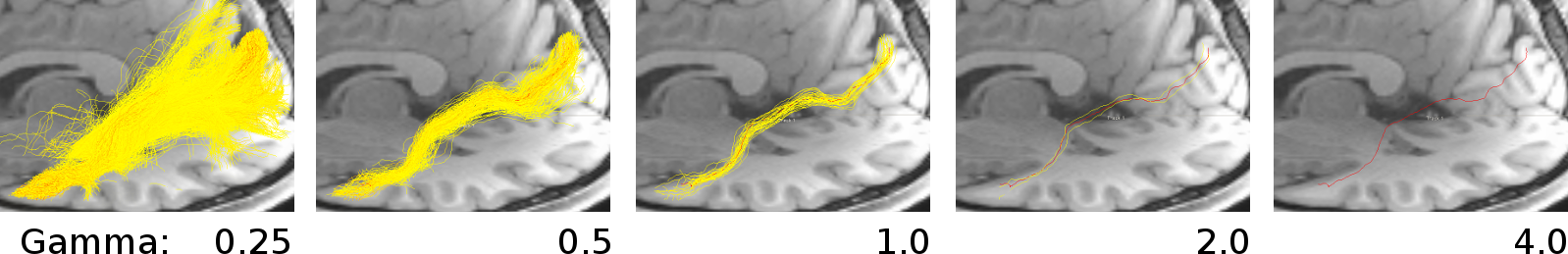}
\caption{We plot a streamline from the inferior longitudinal fasciculus, as well as each tract above $5\%$ of the maximum kernel amplitude for each given $\gamma$. At $\gamma=4$ only the target tract itself is left.}
\end{figure}

In the top row of Fig. \ref{fig:ent} we plot the cross-entropy (surprise) as a function of the number of samples when sampling at random from within the white matter mask. The curve for $\gamma = 0.25$ is nearly flat, while for $\gamma = 0.5$ and $\gamma=1$ the curves quickly bottom out after a short steep learning period. As the length scale increases, the learning rate becomes quite slow. At $\gamma=4$ the curves begin to bottom out at 250000 samples. However, for each of these scales, no significant changes occur after 500000 samples. If there is relevant signal in these samples that is not contained on average in the previous half-million samples, it has a very small relative amount of information about the distribution of tracts.

In general for each length scale both deterministic methods have less entropy than the probabilistic methods. Further, DTI converges (flattens) much faster than the other methods. These results match our prior intuitions about these methods. Probabilistic CSD has the largest information content, though we condition this in that in general noise \emph{adds} information, albeit useless information. Thus, this is not alone a measure of quality, though it does differentiate the method as certainly sampling a tract distribution unlike the others.

In the bottom row of Fig. \ref{fig:ent} we plot the incremental (marginal) entropy rate. For each length scale for each method the marginal information gained per thousand streamlines after the initial learning period is very low; this is unsurprising, but raises questions about the utility of sampling past several million tracts.

These curves are artificially smooth due to the binning of the number of samples by thousands; while in each plot the curves are monotonic, this is an overall trend (at the scale of thousands of samples) and not true locally (at the scale of single samples).

\begin{figure}[t]
\centering
\includegraphics[width=0.45\textwidth]{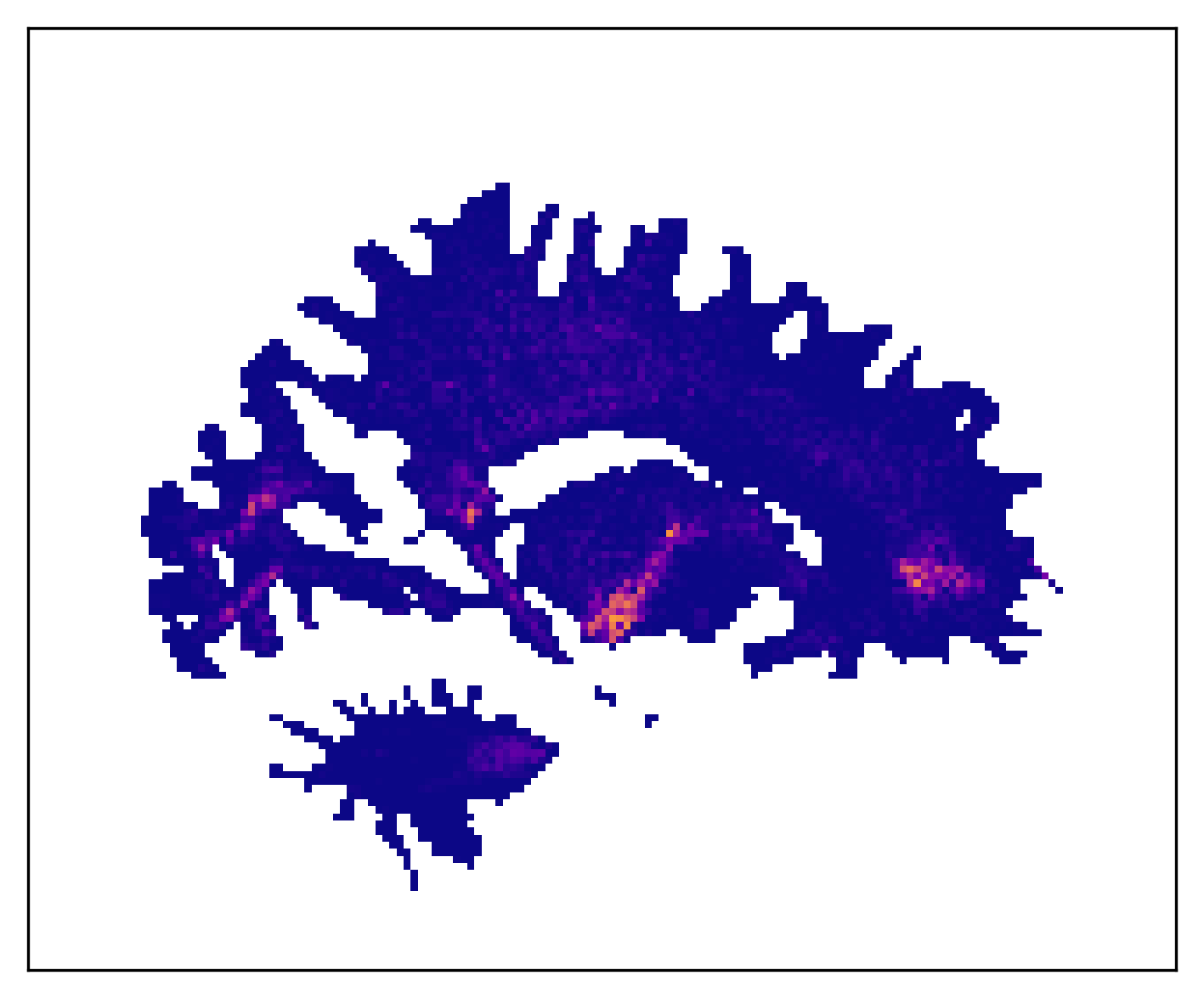}
\includegraphics[width=0.45\textwidth]{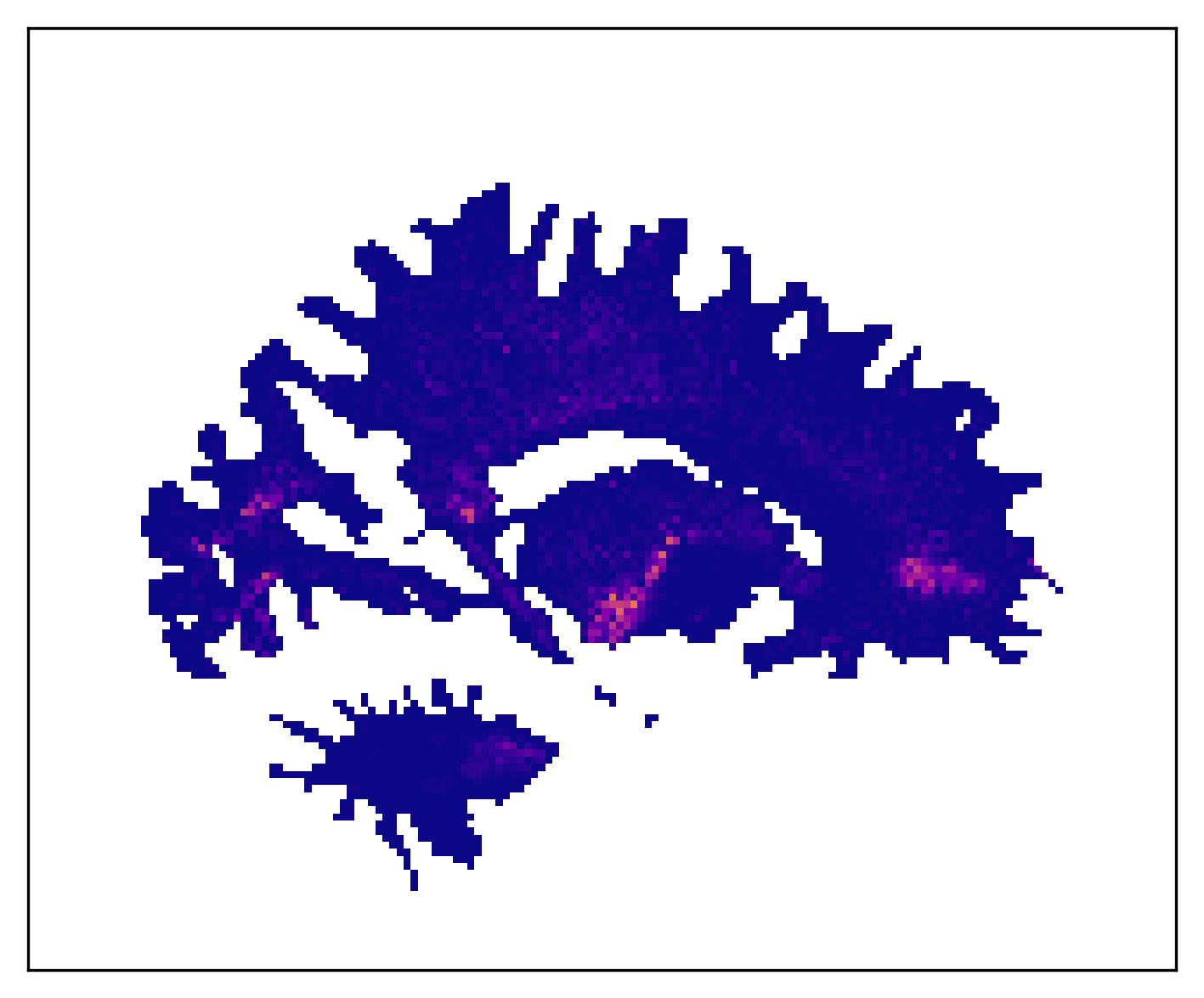}
\includegraphics[width=0.45\textwidth]{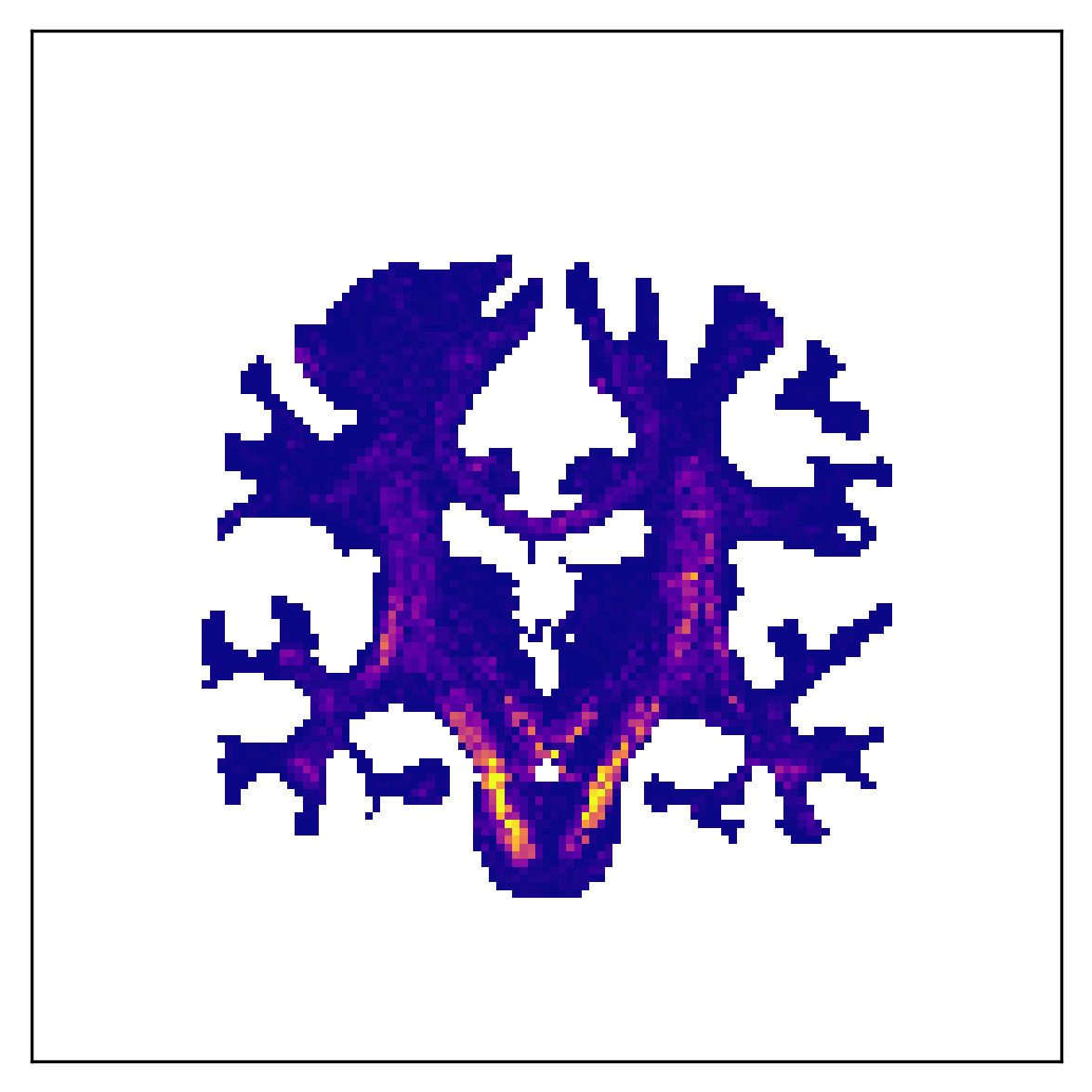}
\includegraphics[width=0.45\textwidth]{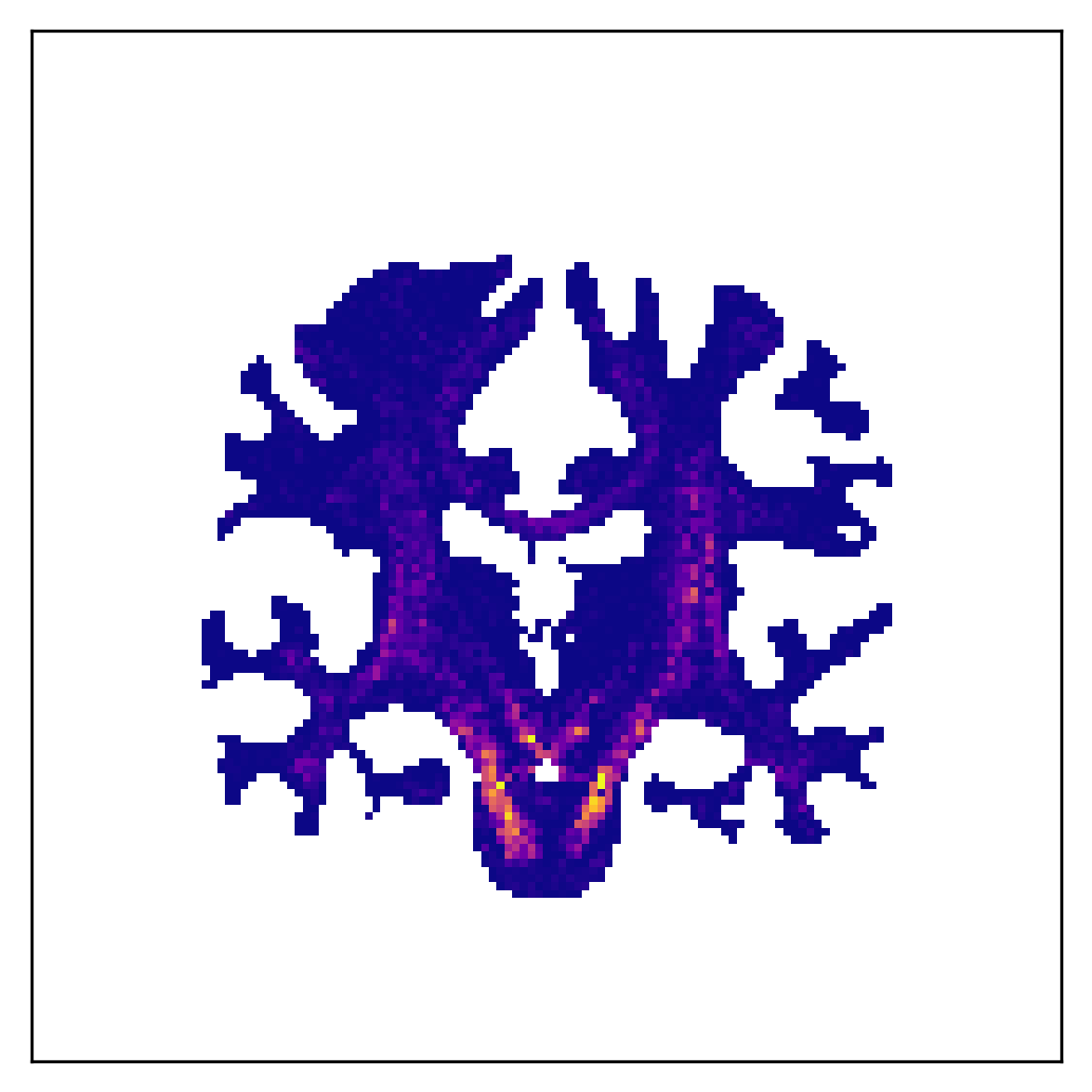}
\includegraphics[width=0.45\textwidth]{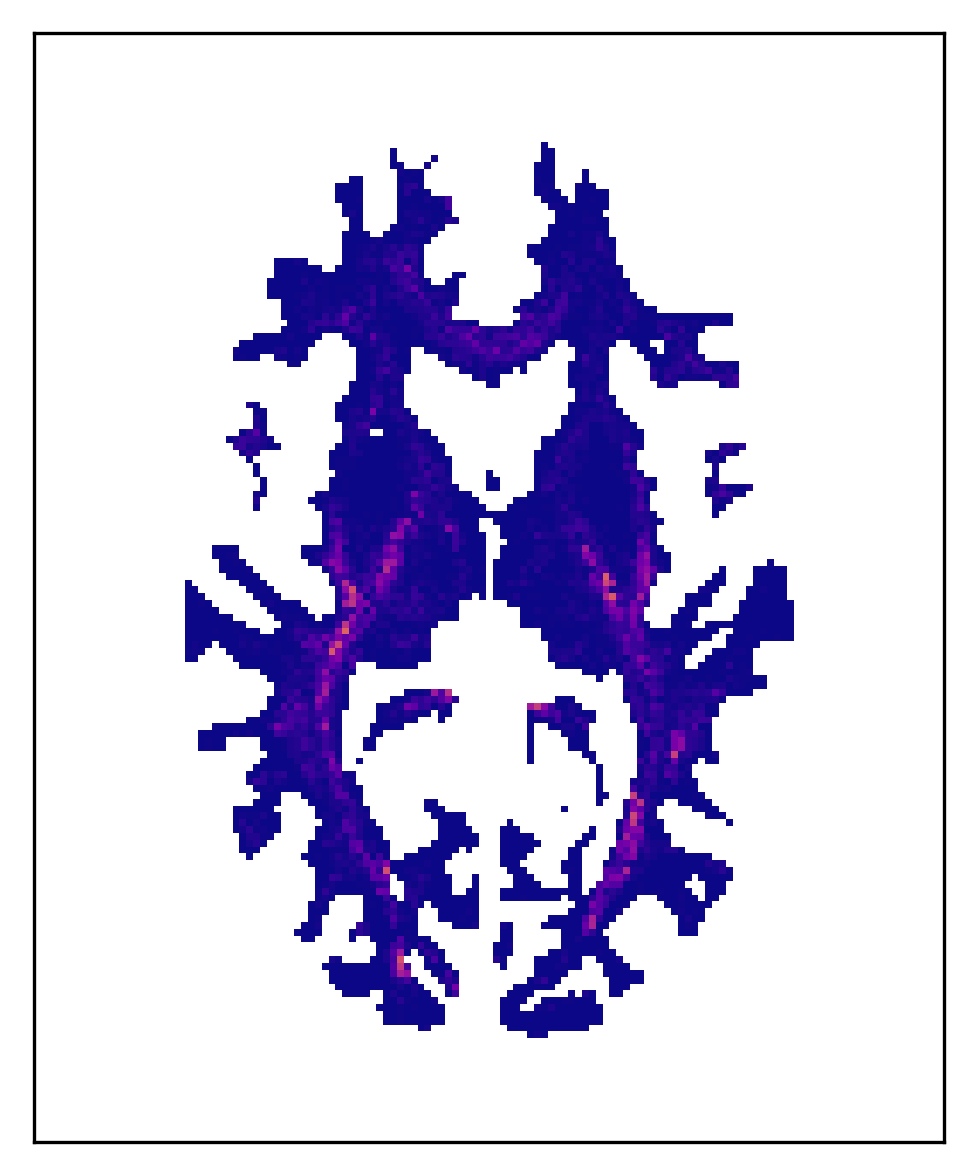}
\includegraphics[width=0.45\textwidth]{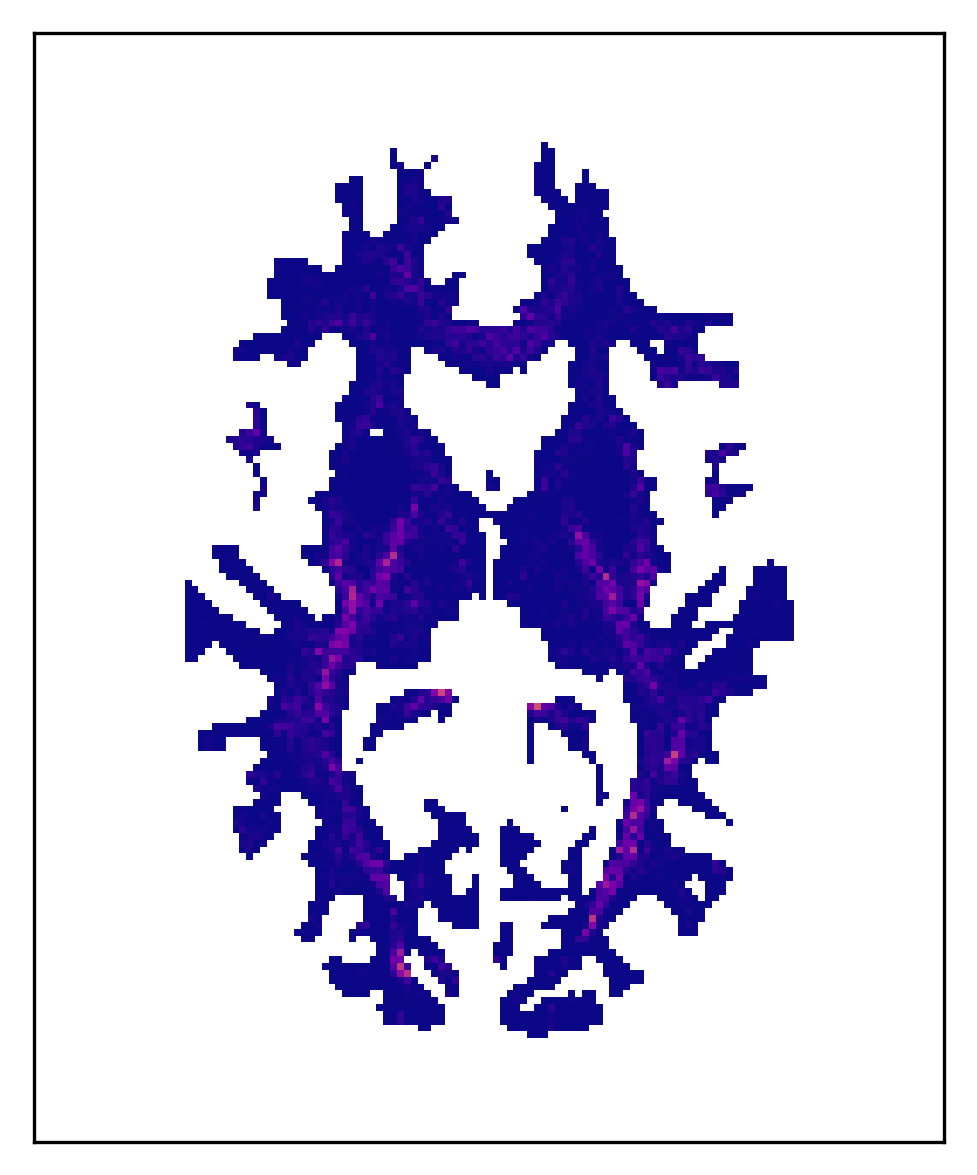}
\includegraphics[width=0.975\textwidth]{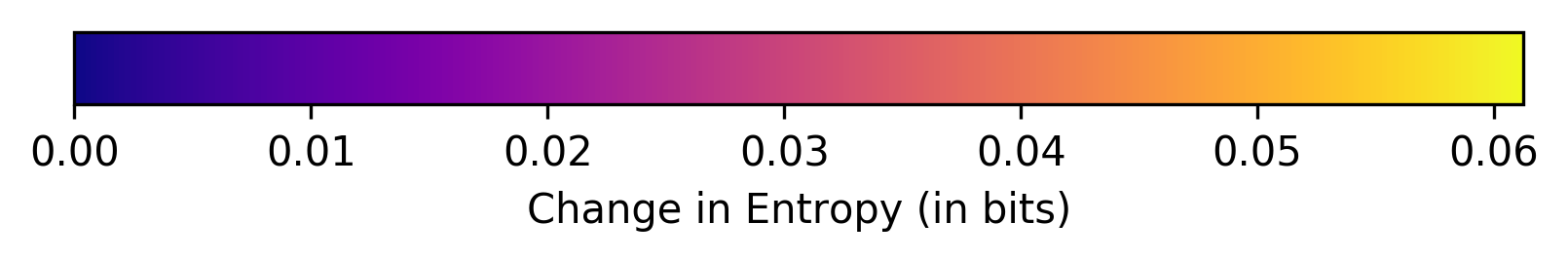}
\label{fig:vx}
\caption{Here are displayed slices of the white matter of one subject, centered on $(60,85,61)$ (MNI). Colors denote the average change in entropy per marginal tract in each voxel, conditioned on seeding in that voxel. The \textbf{left column} shows the average change in the first five tracts seeded in each voxel, while the \textbf{right column} shows the average for tracts 96-100. Most regions converge almost immediately, but specific regions converge very slowly. Note that these measurements \text{do not} account for similarity to tracts seeded in other voxels.}
\end{figure}

We performed a second experiment using the same tracking parameters, but seeding only in a single voxel at a time. For each voxel in the white matter we generated two tractographies (one train set, one test set), seeding only in that voxel 100 times. We then computed the cross-entropy between the two tractographies as a function of the number of samples in the training set. In Figure \ref{fig:vx} we plot the marginal change in entropy over the white matter mask for one subject, both at the start of the process and the end.

For many regions the sampling process converges very quickly. However, for certain regions the sampling process converges quite slowly (albeit without considering samples from other voxels). This suggests at surface level high variance in complexity of the white matter when aggregating into voxels. Further, this provides evidence that a weighted seeding scheme might be more efficient than gridded or uniformly random schema.

\begin{comment}
\begin{figure}[t!]
\begin{picture}(300,225)
\put(0,0){\includegraphics[width=\textwidth]{emp/sigmas_crop.png}}
\end{picture}
\caption{At the \textbf{top} we plot the self cross-entropy (surprise) of each method as a function of the number of tracts, conditioned on the length scale $\gamma$. Below we plot a streamline from the inferior longitudinal fasciculus, as well as each tract above $5\%$ of the maximum kernel amplitude for each given $\gamma$. At $\gamma=4$ only the target tract itself is left.}
\end{figure}
\end{comment}

\section{Discussion}

Our results suggest that in general the number of tracts required to closely approximate the underlying distribution is low using the methods we tested here. This can only be reduced by conditioning tracts or seeds on specific anatomic priors (e.g., seeding only in the corpus callosum or in a specific region). Further, the Human Connectome Project \cite{van2013wu} data are widely regarded as examples of the state of the art. Other sources most likely have less signal content, leading to sharper (more steep) curves and even faster convergence.

On the other hand, this method is inherently limited: we cannot differentiate between ``valid'' signal and spurious systemic error. By sampling to the scales described in this paper researchers can accurately recover the streamline distribution (up to the chosen spatial scale). If this distribution includes ``false'' modes, i.e. modes that not from a corresponding biophysical structure, we \emph{will} consistently recover those false modes. Larger sample sizes in simulations do nothing to guard against false signal; indeed, they reinforce them.

However, spurious random tracts that are not signal (i.e., caused by noise and not artifacts) will have vanishing weight. According to our empirical assessment, this will on average happen rapidly, relative to the weight of the modes.

Consider a tractography method that when queried ten times generates complete noise\footnote{the definition of complete noise is actually tricky, but we could use a proxy of ``drawing a tract length uniformly at random between 1 and the number of voxels, and filling its sequence with points drawn uniformly from the domain of the image''.}
nine times and generates a real tract one time on average. Clearly this method has high noise ($90\%$ useless!) and would not do well in major tractography challenges (e.g., \cite{maier2017challenge}), but, we claim, this hypothetical method would still have value. Since the number of real tracts is finite and the number of possible noise tracts is comparatively large, increasing the sample size will lead to each real tract having more corresponding samples than the noise tracts. Given enough time we may recover the set of real tracts (which we then can simplify to save disk space).

Further consider a tractography method that when queried ten times generates complete noise eight times, generates a real tract one time, and generates a false but non-random tract the remaining time on average. Like the first hypothetical method, this second method produces $90\%$ useless tracts. However, increasing sample size will not lead to vanishing error in this second case due to the false mode (the non-random false tract).

Clearly, no method is purely in the first case; all methods likely have false modes. However, we posit that understanding the number, location, and size of these modes is just as important as understanding a method's information efficiency (i.e., the proportion of true positives for fixed sample size). It is illustrative that F1-score is unable to differentiate between these two hypothetical methods. We believe such considerations should be taken into account, in that valuable methods might be slow to converge, and thus currently ignored.

\section{Conclusion}

In this paper we have investigated the empirical sample complexity of tractography distributions. We have provided a general methodology for estimating and comparing the information content of streamline methods, along with an example analysis of the Human Connectome Project dataset using four different tracking pipelines. 
%We then presented empirical results for four methods showing each to converge for whole-brain tractography within 500,000 tracts.

\subsection*{Acknowledgements}

This work was supported by NIH Grant U54 EB020403 and DARPA grant W911NF-16-1-0575, as well as the NSF Graduate Research Fellowship Program.

%
% ---- Bibliography ----
%

\bibliography{biblio.bib}
\bibliographystyle{splncs03}

\end{document}